\documentclass[preprint]{aastex}
\newcommand{\bbox}[1]{\mbox{\boldmath$#1$}}
\usepackage{graphicx}
\usepackage{psfrag}

\begin{document}

\title{Stokes trapping and planet formation}
\author{M. Wilkinson$^{1}$, B. Mehlig$^{2}$ and V. Uski$^{1}$}
\affil{ $^{1}$Faculty of Mathematics and Computing, The Open
University, Walton Hall,
Milton Keynes, MK7 6AA, England \\
$^{2}$Department of Physics, G\"oteborg University, 41296
Gothenburg, Sweden \\}

\begin{abstract}
It is believed that planets are formed by aggregation of dust
particles suspended in the turbulent gas forming accretion disks
around developing stars. We describe a mechanism, termed \lq
Stokes trapping', by which turbulence limits the growth of
aggregates of dust particles, so that their Stokes number (defined
as the ratio of the damping time of the particles to the
Kolmogorov dissipation timescale) remains close to unity. We
discuss possible mechanisms for avoiding this barrier to further
growth. None of these is found to be satisfactory and we introduce
a new theory which does not involve the growth of small clusters
of dust grains.
\end{abstract}


\maketitle

\section{Background}
\label{sec: 1} It is widely believed that planets are formed by
aggregation of dust particles in an accretion disk surrounding a
growing star; the fact that solar planets have orbits which are
roughly circular and coplanar with the Sun's equator is readily explained
by this model. According to this picture, planet formation is a two-stage
process \citep{Saf69,Arm07}. The final stage must be driven by
gravitational forces, but initially the density of dust particles
is not sufficiently high for gravitational instability to overcome
centrifugal forces. It is believed that the dust particles
initially aggregate by random collisions, and that when the
aggregates have grown to a sufficient size, they sink to the
mid-plane of the accretion disk. Gravitational collapse to form
planets can commence if the density at the mid-plane of the disk
becomes  sufficiently high. In this paper we are concerned with the
first (\lq kinetic') phase of planet formation in this standard model.

The kinetic aggregation process in protoplanetary disks (reviewed
by \citet{Dom07,Hen06,Arm07}) has distinctive features. The dust
particles are surmised to be in a highly turbulent environment (at
least during periods when the star is accreting material at a
substantial rate), because laminar flows could not dissipate
energy sufficiently rapidly to account for observed accretion
rates. Unless some mechanism enables the particles to fuse
together chemically, they must form very weak aggregates (bound by
van der Waals and possibly also electrostatic forces) until
gravitational collapse is well advanced: this implies that the
aggregates are very fragile unless they can be heated to high
temperatures. Also, the aggregates of dust particles may be
fractal structures, with a fractal dimension $D$.

The principal purpose of this paper is to argue that the standard
model for planet formation by aggregation of dust grains is highly problematic,
because the dust grains are unable to aggregate in the turbulent
environment of the accretion disc. Furthermore, we argue that the turbulence
would prevent even very large aggregates from settling to the mid-plane of the disc.
The difficulties are so severe that it is necessary to develop an alternative theory.

A note about terminology is in place here. We describe the
condensed material in the interstellar medium as consisting of
dust {\em grains}, which may form {\em aggregates}, bound by van
der Waals or electrostatic forces. We argue that these aggregates
may {\em cluster} together in space, without coming into contact.
We also describe properties of small fragments of solid material
in a turbulent gas in a more general context, without needing to
specify whether they are dust grains or aggregates: in such cases
we refer to {\em particles} suspended in the gas.

\section{Summary of results}
\label{sec: 2} This paper uses recently acquired insights
\citep{Wil06,Fal02,Wil05} into the dynamics of turbulent aerosols
to draw conclusions about the aggregation of dust particles in a
standard model for a protostellar accretion system. An important
parameter is the Stokes number, ${\rm St}$, which is the ratio of
the time scale for an aggregate of dust grains to be slowed by
drag forces (the \lq stopping time') to the dissipative
correlation time of the turbulent forcing. Note that our
definition of the Stokes number differs from that used in some
other works on planet formation (for example,
\cite{You03,Bau07,You07}), which define the Stokes number as the
ratio of the stopping time to the orbital period, although our
definition is the one which is used in the fluid dynamics
literature, for example in \cite{Fal02,Dun05,Bec06}.

Our principal conclusions may be summarised as follows. We argue
that the relative velocity of the dust aggregates, and therefore
their rate of collision, suddenly increases by several orders of
magnitude when  aggregates grow to a size such that their Stokes
number is of order unity. One consequence of this effect is the
formation of a \lq Stokes trap': aggregates with Stokes numbers
significantly larger than unity will collide with sufficiently
high relative velocity that they fragment upon collision, thereby
replenishing the population of aggregates with smaller Stokes
number. This process would result is a stable population of
aggregates with a Stokes number which has a distribution spanning
${\rm St}\approx 1$. The production of planets therefore depends
upon the particle aggregation process being able to escape from
this Stokes trap. We consider four scenarios by which escape from
the Stokes trap might happen:

{\em 1.~Clustering.} Particles in a turbulent flow with Stokes
numbers of order unity have been shown to cluster together. This
clustering effect has been ascribed to particles being \lq
centrifuged' away from vortices \citep{Max87}. More recent work
has shown that the particles cluster onto a fractal set
\citep{Som93,Bec06,Dun05} of dimension $D_{\rm cl}$. Note that
there are two fractal dimensions in this problem: the dust
particle aggregates may be fractals of dimension $D$, and these
aggregates cluster onto a set of dimension $D_{\rm cl}$. It is
conceivable that this clustering effect might create a means of
escaping the Stokes trap, but we have not been able to identify a
mechanism which would be sufficiently effective in the model which
we have considered.

{\em 2.~Temperature variations.} The temperature of the dust aggregates could
vary dramatically as they are swept back and forth between the
surface and the mid plane of the accretion disk by turbulent
motion of the gas. High temperatures can facilitate chemical
reactions, as can the condensation of water and other volatile
materials when the dust aggregate encounters low temperature
regions. Chemical processes could cause the dust aggregates to
become sufficiently tightly bound that that are less easily
fragmented in collisions. Also, condensation tends to occur more
readily in corners, favouring the formation of compact aggregates.
Below we argue that the accretion disk is sufficiently optically
thick at the crucial stage of its development that there can be a
marked difference between the surface and mid-plane temperatures.
However, unless the star is accreting at a high rate, at distances
significantly greater than $1\,{\rm AU}$ the temperatures are low
everywhere across the profile of the disk. We conclude that
chemical processes are probably not significant unless accretion
is occurring at a very high rate.

{\em 3.~Fractal aggregates.} The Stokes trap can also be avoided
if the dust particles form aggregates with a fractal dimension $D$
less than two: in this case we find that the Stokes number does
not increase as the size of the aggregate increases. Some experiments
on dust grains \citep{Blu00,Kra04} have shown evidence that they
form aggregates with a fractal dimension approximately equal to
$1.4$, others \citep{Wur98} obtained values close to $1.9$. It
should be noted that these experiments only observe the
dust particle aggregates for a short time after their creation,
whereas in protoplanetary processes the aggregates have plenty of
time to relax to a more compact state, which would be
energetically favoured due to increased contact area. The
structure of dust aggregates in real protoplanetary systems is
therefore highly uncertain. The hypothesis that the dust
aggregates are tenuous structures with $D\le 2$ does not appear to
provide a satisfactory resolution, because these tenuous
aggregates will always be advected with the gas, and will not sink
to the midplane. We also show that tenuous aggregates may be torn
apart by shearing forces.

{\em 4.~Quiescence.} It might be expected that the relative
velocities of the dust particle aggregates are reduced if the rate
of accretion is reduced. We write the rate of accretion onto the
star as $\dot M=10^{-7}\Lambda M{\rm yr}^{-1}$, where $M$ is the
solar mass, and the coefficient is chosen so that $\Lambda=1$
corresponds to a typical rate of accretion \citep{Arm07}. We
investigate the scaling of the relative velocity of suspended
particles with the scaling parameter $\Lambda$, and find that it
increases as $\Lambda$ decreases. We conclude that the Stokes trap
is still present in relatively quiescent disks, where $\Lambda$ is
small.

Our calculations are based upon estimates using Kolmogorov's
theory for turbulence \citep{Fri97} in combination with a standard
model for an accretion disk \citep{Sha73}. In our application of
the Kolmogorov theory we assume that the size scale of the largest
eddies, $L$ is proportional to the scale height of the disk, $H$,
writing $L=\ell H$ (we show that this is equivalent to the \lq
$\alpha$-prescription' \citep{Sha73}). We also consider how the
results depend upon the additional scaling parameter $\ell$.

Because of the large uncertainties in the parameters of the model,
our calculations are to be interpreted as illustrations of the
Stokes trapping principle, rather than as quantitative
predictions. Order of magnitude changes in the uncertain parameters will not alter our principle conclusion about
the effectiveness of the Stokes trap in limiting dust aggregation. For simplicity we confine attention to the case where
the heating of the disk is active (that is dominated by
dissipation in the disk rather than by direct illumination from
the star). Our calculations lead to the estimate that typical
particle sizes of Stokes-trapped aggregates at 10 AU from the star
are of the order of $25\mu{\rm m}$.

In principle it is possible to gain information about the sizes of
particles suspended in accretion disks around young stars from
spectroscopic studies, but this is an ill-conditioned problem
\citep{Car04}, with uncertain spectroscopic data. Studies of the
\lq silicate feature' (at approximately $10\mu{\rm m}$ wavelength)
in a survey of spectra of accretion disks around young stars  by
\citet{Kes07} are interpreted as indicating typical particle sizes
between $3$ and $10\mu$m, between $10^{-3}$ AU and $10$AU from the
star. Similar estimates were obtained by \citet{vBo03} and
\citet{Apa05}. By contrast, spectral analysis in the mm-range
(probing predominantly the outer regions of the disk) has been
interpreted as indicating that particles larger than $50\mu{\rm
m}$ may be present \citep{Woo02,Bau07}.

Even if the particle growth process can circumvent the Stokes
trap, the particle aggregates must collapse to the mid-plane of
the disk for gravitational instability to become effective. If we
make the assumption that the integral scale of the turbulence is
comparable to the thickness of the accretion disk (that is, if
$\ell$ is of order one), we show that advection of particles by
the largest turbulent eddies presents another very significant
barrier to planet formation: turbulence will prevent aggregates of
centimetre size from settling to the mid-plane, unless the rate of
accretion is very substantially smaller than $10^{-7}M{\rm
yr}^{-1}$.

We conclude that there is no satisfactory theory for
the formation of planets by aggregation of sub-micron sized dust grains. An alternative theory is therefore required. In the concluding section we introduce an alternative hypothesis.

\section{Collision processes in turbulent aerosols}
\label{sec: 3}
Throughout we assume that the drag force on a particle is proportional to
the difference in velocity between the particle and the
surrounding gas, so that the equation of motion is
\begin{equation}
\label{eq: 3.1} \ddot{\mbox{\boldmath$r$}}=\gamma
(\mbox{\boldmath$u$}(\mbox{\boldmath$r$},t)-\dot{\mbox{\boldmath$r$}})
\end{equation}
where $\mbox{\boldmath$r$}$ is the position of the particle and
$\mbox{\boldmath$u$}(\mbox{\boldmath$r$},t)$ is the fluid velocity
field (until the particles come into contact). This equation is
familiar in the context of Stokes's law for the drag on a sphere,
where the damping rate $\gamma$ is proportional to the kinematic
viscosity $\nu$.

We assume that the gas has a multi-scale turbulent flow, which is
described by the Kolmogorov theory of turbulence \citep{Fri97}. The
length scale and time scale associated with the smallest eddies
where energy is dissipated are denoted by $\eta$ and $\tau$
respectively. According to the Kolmogorov theory of turbulence,
these quantities are determined by the rate of dissipation per
unit mass, ${\cal E}$, and the kinematic viscosity $\nu $:
\begin{equation}
\label{eq: 3.2} \eta = \biggl(\frac{\nu^3}{{\cal E}}\biggr)^{1/4}
\ ,\ \ \ \tau = \biggl(\frac{\nu}{{\cal E}}\biggr)^{1/2}\ .
\end{equation}
If the turbulent motion is driven by forces acting on a
length scale $L\gg \eta$, the velocity fluctuations of the fluid
have a power-law spectrum for wavenumbers between $1/L$ and
$1/\eta$ \citep{Fri97}.

We define the Stokes number to be ${\rm St}=1/(\gamma\tau)$. Very
small particles are advected with the gas flow, but in addition
they have small random velocity due to Brownian diffusion, which
can cause them to collide at a rate ${\cal R}_{\rm d}$. They can
also collide due to the shearing motion of the flow, an effect
discussed by \citet{Saf56} in the context of
rainfall from turbulent clouds. These advective collisions  occur
at a rate ${\cal R}_{\rm a}$ which turns out to be negligible
compared to Brownian diffusion in the present context
(see tables~\ref{tab:2} and \ref{tab:3}).

\begin{figure}[t]
\centerline{\includegraphics[width=14cm]{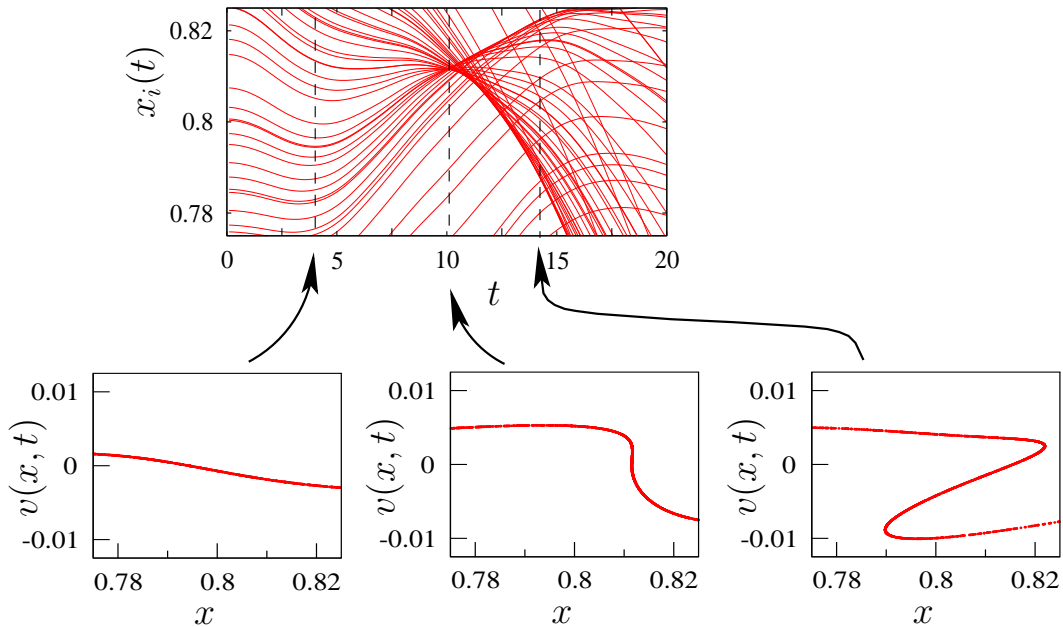}}
\caption{\label{fig: 1} Illustration of fold caustics. Shown are
results from numerical simulations in one spatial dimension of the
equation of motion $\ddot x_i = \gamma (u(x_i,t)-\dot x_i)$, where
$u(x,t)$ is a random velocity field which is a smooth function of
$x$ and $t$. Here $x_i$ denotes the position of particle number
$i$. The top panel shows particle trajectories $x_i(t)$ as a
function of time $t$. The bottom panels show a different
representation of the dynamics: the particle velocities $\dot
x=v(x,t)$ versus their positions $x$ at three different times. The
velocity $v(x,t)$ of the suspended particles is initially
single-valued, but becomes triple-valued in the region between two
fold caustics, which are singularities of the projection of the
phase-space manifold onto coordinate space. In the region between
the caustics the suspended particles have a relative velocity and
their trajectories cross.}
\end{figure}

When the Stokes number approaches unity, there is a dramatic
increase in the relative velocity of suspended particles due to
the formation of fold caustics in their velocity field,
illustrated for the case of one spatial dimension in
figure~\ref{fig: 1}. When faster particles overtake slower ones,
the manifold representing the phase-space distribution of the
particles develops folds, and the velocity field of the particles
goes from being single valued to multi-valued (three-valued, in
this illustration). Before the caustics form the relative velocity
of the particles is due to their Brownian diffusion. After the
folds have formed the relative velocity may be orders of magnitude
higher. The rate of collision ${\cal R}$ for a suspended particle
in a turbulent flow is approximated by a formula presented by
\citet{Meh07}:
\begin{equation}
\label{eq: 3.3} {\cal R}={\cal R}_{\rm d}+{\cal R}_{\rm
a}+\exp(-A/{\rm St}){\cal R}_{\rm g}\ .
\end{equation}
Here ${\cal R}_{\rm g}$ is the collision rate predicted by a model
introduced by \citet{Abr75}, often termed the \lq gas-kinetic model', in which the
suspended particles move with velocities which become uncorrelated
with each other and with the gas flow. A less precise formula was
suggested by \citet{Fal02}, and the idea that there is a dramatic
change in the relative velocity of particles at ${\rm St}\approx 1$ appears
to have been originally proposed by \citet{Mar91}. The exponential term
in (\ref{eq: 3.3}) describes the fraction of the coordinate space for which the
velocity field is multi-valued: $A$ is a \lq universal'
dimensionless constant. The non-analytical dependence of the rate
of caustic production on ${\rm St}$ was noted by \citet{Wil05} and
recent simulations of Navier-Stokes turbulence suggest that
$A\approx 2$ \citep{Pum06}. The rate ${\cal R}_{\rm g}$ greatly
exceeds ${\cal R}_{\rm d}$, but the gas-kinetic theory is only
applicable when the velocity field of the suspended particles is
multi-valued. The exponential term arises because the formation of
caustics is determined by a process involving escape from an
attractor by a diffusion process, similar to the Kramers model for
a chemical reaction. The exponential term is therefore analogous
to the Arrhenius term $\exp(-E/k_{\rm B}T)$ in the expression for
the rate of an activated chemical reaction \citep{Wil05}. The
abrupt increase of the collision rate as the Stokes number exceeds
a threshold value was first noted in numerical experiments by
\citet{Sun97}.

At large Stokes number the rate of collision of particles of
radius $a$ is
\begin{equation}
\label{eq: 3.4} {\cal R}_{\rm g}=4\pi a^2 n \langle \Delta v
\rangle
\end{equation}
where $n$ is the number density of particles and $\Delta v $ is
the relative speed of two suspended particles at the same position
in space (and angular brackets denote averages throughout). In a
multi scale turbulent flow, when $\gamma \tau\ll 1$ the motion of
the suspended particles is underdamped relative to the motion on
the dissipative scale, but it is overdamped relative to slower
long-wavelength motions in the fluid. The relative velocity of two
nearby particles is a result of the different histories of the
particles. If we follow the particles far back in time to when
they had a large separation, their velocities were very different,
but these velocity differences are damped out when the particles
approach each other. \citet{Meh07} surmised the variance of the
relative velocities using the Kolmogorov scaling principle. In the
following we consider, for simplicity, only  the case of a
symmetric collision, where both particles have the same damping
rate, $\gamma$ (in the case of collisions between particles of
very unequal sizes, we may still use the estimate below, using the
smaller of the two values of $\gamma$).

When the particles are underdamped relative to the smallest
dissipative scale, but overdamped relative to the \lq integral'
(driving) scale, we can apply the Kolmogorov cascade principle
\citep{Fri97}, that motion in the inertial range is independent of
the mechanism of dissipation. It is therefore determined by the
rate of dissipation ${\cal E}$ but it does not depend on $\nu$.
The moments of the relative velocity therefore depend only upon
${\cal E}$ and $\gamma$. For the second moment, dimensional
considerations imply that \citep{Meh07}
\begin{equation}
\label{eq: 3.5} \langle \Delta v^2\rangle =K\frac{{\cal
E}}{\gamma}
\end{equation}
where Kolmogorov's 1941 theory of turbulence suggests that $K$ a
universal constant, but in practice $K$ should have a weak
dependence upon Reynolds number due to intermittency effects
\citep{Fri97}. The constant $K$ can be determined by simulation
of Navier-Stokes turbulence, but this has not yet been done.
We use $K=1$ in the remainder of this article.

\citet{Vol80} (see also \citep{Miz88}, \citet{Mar91}) have
discussed relative velocities in the large Stokes
number limit, expressing their results in terms of properties of
the integral velocity and timescales: their results are equivalent
to (\ref{eq: 3.5}), but are harder to apply because the precise
form of the integral scale motion is highly uncertain, whereas the
rate of dissipation in (\ref{eq: 3.5})
is determined directly from the accretion
rate and the mass per unit area in the accretion disk.

\section{Properties of dust-particle aggregates}
\label{sec: 4}

\subsection{Damping rate}
\label{sec: 4.1}
During the kinetic phase of planet formation, the
size $a$ of the suspended particles is usually smaller than the
mean free path of the gas, $\lambda$. In this case the drag force
is proportional to $\rho_{\rm g}c_{\rm s}\bar A (\bbox{u}-\dot{
\mbox{\boldmath$r$}})$, where $\rho_{\rm g}$ is the gas density,
$\bar A$ is the angular average of the projected area of the
particle and $c_{\rm s}$ is the mean molecular speed of the gas
\citep{Eps24}. In this case the damping rate is
\begin{equation}
\label{eq: 4.1} \gamma\sim \frac{\rho_{\rm g}c_{\rm s}\bar A}{m}
\end{equation}
where $m$ is the mass of the particle. In the case of spherical
particles, the coefficient in eq.~(\ref{eq: 4.1}) is $3/4$
\citep{Eps24}.

\subsection{Structure of aggregates}
\label{sec: 4.2}
The suspended particles are expected to consist of dust particles,
of typical size $a_0$, which are weakly bound together by
electrostatic and van der Waals interactions. These aggregates may
have a fractal structure, such that the number of particles in an
aggregate of linear dimension $a$ is
\begin{equation}
\label{eq: 4.2} N\sim \biggl(\frac{a}{a_0}\biggr)^D
\end{equation}
where $D$ is a fractal dimension. The damping rate for an aggregate
of $N$ dust grains with size $a_0$ composed of material of density
$\rho_{\rm p}$ is therefore estimated as
\begin{equation}
\label{eq: 4.3} \gamma \sim \frac{\rho_{\rm g}c_{\rm s}}{\rho_{\rm
p}a_0}N^{\frac{2-D}{D}}=\gamma_0\left(\frac{a}{a_0}\right)^{2-D}\
.
\end{equation}
Since ${\rm St} = 1/(\gamma \tau)$ we have
${\rm St} \propto (a/a_0)^{D-2}$ and conclude that
if $D\le 2$, the Stokes number does not increase as the mass of
the dust aggregate increases.

In the early stages of
aggregation, we expect that the mean free path of the suspended
particles is very large compared to their size, and the relevant
fractal growth mechanism is ballistic aggregation (where particles
incident with random linear trajectories adhere to the aggregate on contact)
rather than diffusion-limited aggregation (where an incoming particle
executes a random walk). Numerical experiments \citep{Nak94} on ballistic
limited aggregation suggest $D\approx 2.98$ for a \lq ballistic
particle cluster aggregate' (BPCA)  model (particles adhering to a
stationary aggregate) and $D\approx 1.93$ for a \lq ballistic
cluster cluster aggregate'  (BCCA) model (where aggregates of
particles move and collide with each other).
(Note that the use of the term \lq cluster' in the designations
of the BPCA and BCCA differs from that used in this paper, as
explained at the end of section \ref{sec: 1}).

\begin{figure}
\includegraphics[width=8cm]{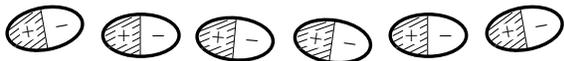}
\caption{\label{fig: 2} Dust grains in a low gravity
environment are found to form aggregates with a low fractal
dimension. This may be due to particles having an electric dipole,
causing them to form chains.}
\end{figure}

However, experiments on agitated dust grains \citep{Blu00,Kra04}
have shown evidence of aggregates with a lower fractal dimension,
$D\approx 1.4$. The lower fractal dimension observed in these
experiments may be related to electrostatic effects: for example
if agitation of dust particles causes them to acquire dipolar
charges, they will tend to form strings (dimension $D=1$), as
illustrated in figure~\ref{fig: 2}. The timescales on which these
aggregates are observed is very short compared to the
astrophysical context which we consider (and also the number of
particles in the aggregates are not very large). The appropriate value for $D$ is,
therefore, highly uncertain. It is probably not universal throughout the
accretion disk structure.

Our interest is
in aggregates which undergo energetic collisions (which are
sufficient to cause fragmentation in some cases), and which have a
long time to relax into a compact configuration with lowest
energy. In the following we therefore usually assume that the
aggregates are predominantly in compact configurations, with
$D\approx 3$, but in many cases below we continue to give formulae
for general values of $D$.

\subsection{Strength of aggregates}
\label{sec: 4.3}

The other distinctive aspect of the dust-grain aggregates concerns
their fragility under collisions. There is a substantial
literature on this topic: see \citet{Dom97}, \citet{You03} and
references therein. In the following we introduce a new treatment,
valid in the limit where the grains are very small.

We mentioned earlier that the aggregates are bound together by van
der Waals forces, which means that they must be very fragile. For
a collision between two aggregates of size $a_1$ and $a_2$, there
will be a velocity $v_{\rm cr}$ above which the average change in
the mass of the larger particle is negative. It is extremely hard
to estimate this critical velocity for collisions of aggregates.
We estimate this critical velocity below for the collision of two
grains. We will assume that the critical velocity is approximately
independent of size, so that the two-grain estimate is sufficient
for the general case. In the following we therefore use $v_{\rm
cr}$ to refer to the mean value of the relative velocity below
which {\em pairs} of grains remain bound.

We motivate this assumption by the following argument. When two
aggregates collide, initially the collision only influences the
grains on the surface of the aggregates in the vicinity of the
point of collision. If the aggregates collide with a relative
velocity $v$, some of the grains will acquire relative velocities
which are comparable to $v$. If $v/v_{\rm cr}$ is significantly
greater than unity, grains on the periphery of the colliding
aggregates will be able to escape.

In order to estimate the critical relative
collision velocity above which fragmentation occurs we therefore
estimate the critical velocity $v_{\rm cr}$ above which two dust grains do not remain
bound after a binary collision. This will be estimated by equating the kinetic
energy of the collision process between two dust grains, treated as spheres
of radius $a_0$, to the binding energy.

The binding energy is usually estimated by a theory proposed by
\citet{Cho93} (and discussed in simpler terms by \citet{You03}),
which assumes that the spheres deform so that they are in contact
on a small circular patch. By balancing the force required to
deform the spheres against a derivative of the surface energy
released by bringing them into contact, the equilibrium
deformation and hence the binding energy are estimated in terms of
a Young's modulus for the spheres, $Y$, and their surface energy
of attraction, $\sigma$. Omitting the dimensionless prefactors
(which depend upon making more precise definitions of $Y$,
$\sigma$) the surface deformation $\delta$ and binding energy
$E_{\rm b}$ are estimated to be
\begin{equation}
\label{eq: 4.4}
\delta\sim \left(\frac{\sigma^2a_0}{Y^2}\right)^{1/3}   \ ,\ \ \
E_{\rm b}\sim \left(\frac{\sigma^5a_0^4}{Y^2}\right)^{1/3}\ .
\end{equation}
The latter expression gives an estimate for the critical velocity $v_{\rm cr}$,
obtained in \citet{Cho93}:
\begin{equation}
\label{eq: 4.5}
v_{\rm cr}\sim\left(\frac{\sigma^5}{\rho_{\rm p}^3Y^2a_0^5}\right)^{1/6}\ .
\end{equation}
Using values for quartz of $Y=5.4\times 10^{10}\,{\rm N\,m}^{-2}$,
$\sigma=2.5\times 10^{-2}\,{\rm N\,m}^{-1}$, $\rho_{\rm
p}=2.6\times 10^3\,{\rm kg\,m}^{-3}$ quoted in \cite{Cho93}, we
estimate $\delta\approx 3\times 10^{-11}{\rm m}$ and $v_{\rm
cr}\approx 0.2\,{\rm m\,s}^{-1}$ for grains of size
$a_0=10^{-7}{\rm m}$. The validity of this approach is
questionable when the value of the deformation is small compared
to the size of an atom, as is found in this example. We therefore
propose an alternative estimate of the binding energy.

\begin{figure}[t]
{\includegraphics[width=5cm]{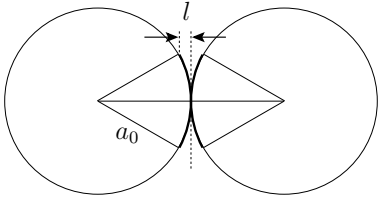}} \caption{\label{fig:4} Two
spheres in contact have a surface area $\pi l a_0$ with separation
less than $2l$.}
\end{figure}

We assume that there is a van der Waals binding energy $\delta E_1$ between
atoms, which operates over a range $l$, which we equate with the typical interatomic
distance. We assume that there is a binding energy $\delta E_1/l^2$ per unit
area for all parts of two spheres which are within a distance
$2l\ll a_0$ of each other, where $\delta E$ is the van der Waals
binding energy between two atoms, and $l$ is the range over which
it acts. Two spheres in contact have a surface area $\pi l a_0$
with separation less than $2l$ (see figure \ref{fig:4}), so that
the binding energy is $E_{\rm b}\sim \pi \delta E a_0/l$. Unless
the grains are made of very plastic material, particles are not
expected to aggregate if their (centre of mass frame) kinetic
energy is large compared to the binding energy. Writing $E_{\rm
b}\sim m v_{\rm cr}^2\sim \rho_{\rm p}a_0^3 v_{\rm cr}^2$, we
estimate a critical velocity
\begin{equation}
\label{eq: 4.6}
v_{\rm cr}=\sqrt{\delta E_1 /\rho_{\rm
p}l a_0^2}\ .
\end{equation}
Taking $\delta E_1\approx 10^{-21}{\rm J}$, $l=3\times
10^{-10}{\rm m}$ and $a_0=10^{-7}{\rm m}$, we find that $v_{\rm
cr}\approx 3\times 10^{-1}{\rm m\,s}^{-1}$ for binding of
microscopic grains. Thus collisions between aggregates with
relative velocities of the order of $30\,{\rm cms}^{-1}$ have the
potential to dislodge particles from an aggregate.

It is instructive to compare the microscopic and the macroscopic
models, equations (\ref{eq: 4.5}) and (\ref{eq: 4.6}). The surface
energy and Young's modulus can be related to microscopic
parameters by writing $\sigma\sim \delta E_1/l^2$, where $\delta
E_1\approx 10^{-21}\,{\rm J}$ is the van der Waals interaction
energy and $Y\sim \delta E_2/l^3$, where $\delta E_2\approx
10^{-18}\,{\rm J}$ is the characteristic energy scale for covalent
bonds: these expressions give values which are consistent with the
values for $\sigma $ and $Y$ for quartz quoted above. Expressing
equation (\ref{eq: 4.5}) in terms of microscopic quantities gives
\begin{equation}
\label{eq: 4.7}
v_{\rm cr}\sim \sqrt{\delta E_1 /\rho_{\rm
p}l a_0^2}\left(\frac{\delta E_1}{\delta E_2}\right)^{1/3}\left(\frac{a_0}{l}\right)^{1/6}\ .
\end{equation}
We see that the two estimates (\ref{eq: 4.5}) and (\ref{eq: 4.6})
differ by dimensionless ratios with small exponents, and the
values obtained from the two formulae are therefore typically
quite similar. Whichever of equation (\ref{eq: 4.6}) or (\ref{eq:
4.5}) gives the larger value should be used to estimate the
critical velocity. The material parameters can make a significant
difference, in particular ice has a much larger surface energy and
is softer than quartz, and the predicted critical velocity is
approximately an order of magnitude larger for ice \citep{Cho93}.

There are significant sources of uncertainty in these estimates
for the critical velocity for binding single grains, which are
very hard to quantify. One is the effect of surface roughness.
This will reduce the effective contact area between two grains,
and therefore reduce $v_{\rm cr}$. The effect of electrostatic
charging of dust grains and aggregates is harder to quantify. It
is possible that the microscopic grains will all tend to acquire a
positive charge, due to ionisation by energetic photons. This
would reduce the binding energy. Another possibility is that
grains acquire random charges through frictional charge transfer
in collisions. This may increase the critical velocity for pairs
of grains, because of electrostatic attraction of oppositely
charged particles. This effect will be reduced by cancellation of
charges for larger aggregates.

Experiments to determine the critical velocity for collisions of grains have been
performed by projecting small silica spheres at a glass surface \citep{Pop97}. These experiments
have usually showed that the spheres adhere to the surface at velocities which are approximately
ten times higher than those predicted by equation (\ref{eq: 4.5}). The reasons for this
discrepancy are not understood. Although the effects of charges on the particles were quantified
 in these experiments, it is possible that effects of electrostatic charges on the surface
might account for the discrepancy.

\section{Turbulence in protoplanetary disks}
\label{sec: 5}

We now consider estimates relating to dust
particles suspended in gas surrounding a growing star. Observational
evidence suggests that there
is usually sufficient angular momentum that the gas forms an
accretion disk and we use the steady-state disk model described
by \citet{Sha73}. The kinetic and potential energy of the material
in this cloud must be dissipated into heat and radiated away in
order to allow the material to fall into the growing star: we
consider the case where the disk is primarily heated by
dissipation, rather than by radiation from the star. It is known
that stars can form quite rapidly (over a timescale of $t_{\rm
acc}\approx 10^6{\rm yr}$). The process
is so fast that either shocks or turbulent processes must play a
role in the dissipation of energy: the rate of dissipation by
laminar flow is too small by many orders of magnitude. It has been
argued that magneto-hydrodynamic mechanisms provide the necessary
large-scale instability \citep{Arm07}, but the gas is only weakly
ionised and the small scale dissipation is expected to be
described by conventional hydrodynamics, for which the Kolmogorov
theory of turbulence is applicable. Shocks dissipate a finite
fraction of the kinetic energy of a gas almost instantaneously, so
that their role must be short lived and throughout most of the
accretion process turbulence is the dissipation mechanism.

In the following we consider a star of mass $M$ (taken to be one
solar mass) surrounded by a protoplanetary nebula, which is
collapsing into the star at a rate $\dot M$. We assume
\begin{equation}\label{eq: 5.0}
\dot M=10^{-7}\Lambda M{\rm yr}^{-1}\ .
\end{equation}
Most discussions of protoplanetary accretion systems assume that
$\Lambda \approx 1$. We assume that the accretion zone is
predominantly molecular hydrogen (we take the mean molecular mass
of the gas molecules $m_{\rm g}=\mu m_{\rm H}$, with $\mu=2.34$
and $m_{\rm H}=1.67\times 10^{-27}{\rm kg}$ being the mass of
atomic hydrogen). For estimating the mean-free path of the gas, we
assume that the molecular collision cross section is $S_{\rm
col}=2\times 10^{-19}{\rm m}^2$. These and other assumed parameter
values are collected in table~\ref{tab:tab0}. We use
\begin{equation}
\label{eq: 5.1}c_{\rm s}^2=k_{\rm B}T/m_{\rm g}
\end{equation}
to relate $c_{\rm s}$, a characteristic speed which is of the
order of the speed of sound, to temperature.

\begin{deluxetable}{lcc}
\tablecolumns{3} \footnotesize \tablewidth{14cm}
\tablecaption{\label{tab:tab0} Assumed values for parameters.}
\tablehead{\colhead{Variable}&\colhead{Symbol}&\colhead{Value}}
\startdata Mass of gas molecules &$m_{\rm g} = \mu m_{\rm H}$ &$
\mu = 2.34   $              \nl Molecular collision cross section
&$S_{\rm col}$& $2\times 10^{-19}{\rm m}^2$     \nl Mass of the
star &$M$      & $1.99 \times 10^{30}{\rm kg}$
\nl Typical size of interstellar dust grains & $a_0$&
$10^{-7}{\rm m}$           \nl Density  of suspended particles
& $\rho_{\rm p}$& $2\times 10^3{\rm kg}/{\rm m}^3$\nl Mass
fraction of suspended particles to gas molecules&$\kappa$& $3/170$
\nl
\enddata
\end{deluxetable}

The accretion disk is characterised by its mass density per unit
area $\Sigma(R)$ at a distance $R$ from the star, its scale height
$H(R)$, its rate of dissipation per unit area $Q(R)$, and its
surface temperature $T(R)$. Using the continuity equations for
flow of mass and angular momentum within a thin accretion disk
which is in a quasi-stationary state, the rate of dissipation per
unit area $Q$ at radius $R$ can be obtained in terms of the rate
of accretion, $\dot M$ \citep{Sha73}:
\begin{equation}
\label{eq: 5.2} Q(R)=\frac{3}{8\pi}\frac{GM\dot M}
{R^3}\biggl[1-\biggl(\frac{R_{\rm c}}{R}\biggr)^{1/2}\biggr]\,.
\end{equation}
Here $R_{\rm c}$ is the radius of the core of the accretion
system. In the following we concentrate on the region $R/R_{\rm
c}\gg 1$, and do not carry forward the final factor of the above
expression. Note that this formula is independent of the actual
mechanism of dissipation. The rate of dissipation per unit mass is
obtained from the mass per unit area $\Sigma(R)$ of the accretion
disk at radius $R$, and the density of gas $\rho_g(R)$ at the
mid-plane of the disk is related to the scale height:
\begin{equation}
\label{eq: 5.3} {\cal E}(R)=\frac{Q(R)}{\Sigma(R)}\ ,\ \ \
\rho_{\rm g}(R)=\frac{\Sigma(R)}{\sqrt{2\pi}H(R)}
\end{equation}
where in the second expression we assume a Gaussian density
profile with variance $H$. Consider next how to estimate the
height of the disk, $H(R)$. The gas is in Maxwell-Boltzmann
equilibrium in the gravitational potential and we assume that the
contribution to the potential from the disk itself is negligible.
The potential energy at a distance $z$ from the mid plane is
$\Phi(z,R)=GMm_{\rm g}z^2/2R^3$, so that the gas density at
distance $z$ from the mid plane of the disk is
$\rho(z,R)=\rho_{\rm g}(R)\exp[-z^2/2H^2(R)]$, the scale height of
the disk being
\begin{equation}
\label{eq: 5.4} H(R)=\biggl({kT(R)R^3\over{GMm_{\rm
g}}}\biggr)^{1/2}=c_{\rm s}(R)/\Omega(R)
\end{equation}
where $\Omega(R)=\sqrt{GM/R^3}$ is the Keplerian orbital angular
frequency for a mass $M$ at radius $R$.

We assume that the surface temperature $T(R)$ is determined by the
rate at which the thermal energy created by dissipation of
turbulent motion can be radiated away. This assumption is
justifiable as long as $\Lambda$ in equation (\ref{eq: 5.0}) is
not too small, but when $\Lambda\ll 1$ the rate of heating by
radiation from the star is expected to be significant
\citep{Arm07}. Assuming that the surface of the disk behaves as a
black body with emissivity $\varepsilon$, one obtains
\begin{equation}
\label{eq: 5.5} Q(R)=2\varepsilon \sigma T^4(R)\sim
{3\over{8\pi}}{GM\dot M\over{R^3}}
\end{equation}
where $\sigma$ is the Stefan-Boltzmann constant. Assuming
$\varepsilon \approx 1$, one obtains the radial temperature
profile
\begin{equation}
\label{eq: 5.6} T(R)=\left(\frac{3\Omega^2\dot M}{16\pi
\sigma}\right)^{1/4}=T(R_0)\Lambda^{1/4}\left(\frac{R}{R_0}\right)^{-3/4}\
.
\end{equation}
where $R_0$ is a convenient reference radius, which we take to be
$R_0=1\,{\rm AU}=1.5\times 10^{11}\,{\rm m}$, and $T(R_0)$ is the
temperature at $R_0$ with $\Lambda$ set equal to unity. Given the
radial dependence $T(R)$ of the temperature, the radial dependence
of the velocity of sound and of the disk thickness can be
determined. Using equations (\ref{eq: 5.5}) and (\ref{eq: 5.4}) we
find
\begin{equation}
\label{eq: 5.7}c_{\rm s}(R)=c_{\rm
s}(R_0)\Lambda^{1/8}\biggl({R\over{R_0}}\biggr)^{-3/8}
\end{equation}
\begin{equation}
\label{eq: 5.7a}
H(R)=H(R_0)\Lambda^{1/8}\biggl({R\over{R_0}}\biggr)^{9/8}\ .
\end{equation}
(Again, the dependence on $\Lambda$ is not shown explicitly in the
list of arguments.) In order to estimate the radial dependence
of other quantities it is necessary to determine the radial
dependence of the density. This is achieved as follows. Applying
the continuity equation for angular momentum leads to a relation
between $\Sigma$ and an effective kinematic viscosity (termed the
eddy viscosity), $\nu_{\rm eff}$, that is the diffusion
coefficient characterising the transport of angular momentum
across the accretion disk \citep{Sha73}:
\begin{equation}
\label{eq: 5.8} \dot M=3\pi \nu_{\rm eff}(R)\Sigma(R)\ .
\end{equation}
The eddy viscosity in turn is determined by the size $L$ of the largest eddies
in the turbulent flow (the integral scale).
We estimate $\nu_{\rm eff}\sim
L^2/t_L$, where $t_L$ is the eddy turnover time for an eddy of
size $L$. The Kolmogorov scaling argument implies that $t_L\sim
(L^2/{\cal E})^{1/3}$, so that we estimate
\begin{equation}
\label{eq: 5.9} \nu_{\rm eff}\sim
L^{4/3}Q^{1/3}(R)\Sigma^{-1/3}(R)\ .
\end{equation}
It is natural to assume that $L$ is of the order of $H(R)$ and we
write
\begin{equation}
\label{eq: 5.9a} L = \ell\, H(R)
\end{equation}
where $\ell$ is a parameter. Taking (\ref{eq: 5.4}), (\ref{eq:
5.5}),(\ref{eq: 5.8}) and (\ref{eq: 5.9}) together, we find
\begin{equation}
\label{eq: 5.10} \Sigma(R) = \frac{2\sqrt{2}}{9\pi
\ell^2}\frac{\dot M}{H^2\Omega} .
\end{equation}
Usually it is assumed \citep{Sha73} that $\nu_{\rm eff}\sim \alpha
c_{\rm s}H$ (where $\alpha < 1$ is a dimensionless coefficient).
Our own approach is equivalent to this \lq $\alpha$-prescription':
using (\ref{eq: 5.10}) in equation (\ref{eq: 5.8}), we obtain
\begin{equation}
\label{eq: 5.10a} \nu_{\rm eff}= \frac{3}{2\sqrt{2}}\ell^2 c_{\rm
s}H\ .
\end{equation}
Thus we see that the $\alpha$-presription is equivalent to
assuming that the integral scale of the turbulence is smaller than
$H$ by a factor $\ell \approx \sqrt{\alpha}$. It is widely
accepted that may many observations are compatible with
$\alpha\approx 10^{-2}$ \citep{Har98}, corresponding to $L\approx
H/10$. Accordingly, wherever we quote numerical values for
quantities without specifying how they scale with $\ell$, we have
set $\ell=0.1$.
An advantage of our \lq $\ell$-prescription' is that it makes the physical nature of the
adjustable parameter clearer than for the standard \lq $\alpha$-prescription'. The
rather small values of $\alpha$ indicated by observations could be indicative of a fundamental
problem with a theory. Our alternative approach is reassuring because it indicates that a
more physically transparent parameter, $\ell=L/H$, is not in fact a very small number.

These results allow us to determine the power-law radial
dependence of other quantities. The results for those determining
the gaseous component of the accretion disk are summarised in
table~\ref{tab:tab1}, writing a generic variable in the form
\begin{equation}
\label{eq: 5.11}
X=X(R_0)\left(\frac{R}{R_0}\right)^{\delta_R}\Lambda^{\delta_\Lambda}\ell^{\delta_\ell}\,.
\end{equation}
where $R_0=1\,{\rm AU}=1.5\times 10^{11}\,{\rm m}$, and $X(R_0)$
is the value of $X$ at radius $R_0$ with $\Lambda=\ell=1$.

\begin{deluxetable}{lcccccc}
\tablecolumns{7} \footnotesize \tablewidth{16cm} \tablecaption{
\label{tab:tab1} Estimates of quantities characterising the
gaseous part of the accretion disk, quoted to two significant
figures.}
\tablehead{\colhead{Variable}&\colhead{Symbol}&\colhead{Equation}&\colhead{$X(R_0)$}&\colhead{$\delta_R$}&
           \colhead{$\delta_\Lambda$}& \colhead{$\delta_\ell$}}
           \startdata
Surface temperature&$T$&  eq.(\ref{eq: 5.6})&$130\,{\rm K}$&$-3/4$&$1/4$&$0$   \\
Speed of sound&$c_{\rm s}$ & eq.(\ref{eq: 5.7})&$670\,{\rm m\,s}^{-1}$&$-3/8$&$1/8$&$0$   \\
Disk height&$H$& eq.(\ref{eq: 5.7a})&$3.4\times10^9\,{\rm m}$&$9/8$&$1/8$&$0$   \\
Surface density&$\Sigma$&eq.(\ref{eq: 5.10})  & $280\,{\rm kg\,m}^{-2}$& $-3/4$&$3/4$&$-2$   \\
Gas density&$\rho_{\rm g}$&eq.(\ref{eq: 5.3})&$3.3\times 10^{-6}\,{\rm kg\,m}^{-3}$& $-15/8$&$5/8$&$-2$   \\
Dissipation rate&${\cal E}$&eq.(\ref{eq: 5.3})&$0.11\,{\rm m}^2{\rm s}^{-3}$&$-9/4$&$1/4$&$2$   \\
Gas mean-free path&$\lambda$&$\mu m_{\rm H}/(\sqrt{2}\rho_{\rm g}S)$&$0.42\,{\rm m}$& $15/8$&$-5/8$&$2$   \\
Kinematic viscosity&$\nu$ & $\nu=\lambda c_{\rm s}$ & $280\,{\rm m}^2{\rm s}^{-1}$&$3/2$&$-1/2$&$2$   \\
Kolmogorov length&$\eta$ & eq.(\ref{eq: 3.2}) & $120\,{\rm m}$ & $27/16$&$-7/16$&$1$   \\
Kolmogorov time &$\tau$ & eq.(\ref{eq: 3.2}) & $52\,{\rm s}$ & $15/8$&$-3/8$&$0$   \\
Kolmogorov velocity&$u_{\rm K}$ & $\eta/\tau$ & $2.3\, {\rm m\,s}^{-1}$ & $-3/16$&$-1/16$&$1$   \\
Integral velocity&$u_L$ & $u_L=({\cal E}L)^{1/3}$ & $710\, {\rm m\,s}^{-1}$ & $-3/8$&$1/8$&$1$   \\
Integral timescale&$t_L$ & $t_L=L/u_L$ & $4.6\times 10^6\, {\rm s}$ & $3/2$&$0$&$0$   \\
\enddata
\end{deluxetable}

\section{Relative velocities, collision rates and Stokes trapping}
\label{sec: 6}

Next we consider the behaviour of dust grains suspended in the gas
forming the accretion disk. We consider the case where the dust
grains are almost all sub-micron sized particles: we assume they
are composed of material with density $\rho_{\rm p} = 2\times
10^3{\rm kgm}^{-3}$, and that these are initially spherical
particles of radius $a_0=10^{-7}{\rm m}$. Following \citet{Hay81},
we assume that the mass ratio $\kappa$ of suspended particles to
gas molecules is $\kappa =3/170\approx 0.018$.

The results described above can be used to estimate collision
rates. We consider two cases. First, we estimate the collision
rate when the Stokes number is small, and when the collision
mechanism is Brownian diffusion (we shall see that advective
collisions \citep{Saf56} make a negligible contribution). Second,
we also require the rate of collision for large Stokes number.
Both estimates require the number density of dust grains. We write
\begin{equation}
\label{eq: 6.1} n\sim {\kappa \rho_{\rm
g}\over{m_0}}\left(\frac{a}{a_0}\right)^{-D}=n_0\left(\frac{a}{a_0}\right)^{-D}
\end{equation}
where $D$ is a fractal dimension and $\kappa\approx 0.018$ is the
ratio of the mass densities of condensed to gaseous matter and
$m_0=4\pi\rho_{\rm p}a_0^3/3$ is the mass of a microscopic dust
grains; the second equality defines $n_0$.

The mean speed of particles due to Brownian motion is
\begin{equation}
\label{eq: 6.2} \langle v_{\rm d}\rangle\sim c_{\rm s}\sqrt{\mu
m_{\rm H}\over{{N
m_0}}}=v_0\left(\frac{a}{a_0}\right)^{-D/2}\nonumber
\end{equation}
where the second equality defines $v_0$. Our estimate of the collision rate
due to Brownian diffusion for a general value of the fractal dimension of
the aggregates is
\begin{eqnarray}
\label{eq: 6.3}
\nonumber       {\cal R}_{\rm d}&\sim& 4\pi\sqrt{2}na^2\langle
v_{\rm d}\rangle\sim
4\pi\sqrt{2}n_0a_0^2v_0\left(\frac{a}{a_0}\right)^{4-3D\over
2}\\&=&{\cal R}_{{\rm d}0}\left(\frac{a}{a_0}\right)^{4-3D\over 2}\
\end{eqnarray}
and the factor ${\cal R}_{{\rm d}0}$ is given in table \ref{tab:2}.
The collision rate due to advective shearing is approximately
\begin{equation}
\label{eq: 6.8} {\cal R}_{\rm a}\sim \frac{na^3}{\tau}\ .
\end{equation}
The gas-kinetic collision rate (taking $K=1$ in equation (\ref{eq: 3.5})) is
\begin{eqnarray}
\label{eq: 6.4}
\nonumber       {\cal R}_{\rm g}&=&4\pi n a^2\sqrt{\langle \Delta
v^2\rangle}=4\pi n_0a_0^2\sqrt{{\cal
E}\over{\gamma_0}}\left(\frac{a}{a_0}\right)^{2-D\over 2}\\&=&{\cal
R}_{{\rm g}0}\left(\frac{a}{a_0}\right)^{2-D\over 2}\ .
\end{eqnarray}
The Stokes number is
\begin{equation}
\label{eq: 6.5} {\rm St}={1\over{\gamma \tau}}={\rm
St}_0\left(\frac{a}{a_0}\right)^{D-2}\ .
\end{equation}
All of these quantities have a power-law dependence upon radius
and particle size: for the generic quantity $X$ we write
\begin{equation}
\label{eq: 6.6} X =X(R_0,a_0)\left(\frac{R}{R_0}\right)^{\delta_R}
\Lambda^{\delta_\Lambda}\,
\ell^{\delta_\ell}\left(\frac{a}{a_0}\right)^{\delta_a}
\end{equation}
(again, it is to be understood that $X(R_0,a_0)$ is evaluated for
$\Lambda=\ell=1$). The quantities determining the collision rates
are collected in table~\ref{tab:2}.

\begin{deluxetable}{lccccccc}
\rotate \footnotesize \tablewidth{18.5cm} \tablecolumns{7}
\tablecaption{\label{tab:2} Estimates of quantities characterising
the suspended particles (quoted to two significant figures).}
\tablehead{
Variable&Symbol&Equation&$X(R_0,a_0)$&$\delta_R$&$\delta_\Lambda$&$\delta_\ell$&$\delta_a$}
\startdata
Damping rate&$\gamma$&eq.(\ref{eq: 4.3})&$0.11\,{\rm s}^{-1}$&$-9/4$&$3/4$&$-2$&$2-D$\\
Number density&$n$ &eq.(\ref{eq: 6.1})&$7.0\times 10^{7}\,{\rm m}^{-3}$&$-15/8$&$5/8$&$-2$&$-D$\\
Diffusive velocity&$v_d$&eq.(\ref{eq: 6.2})&$1.4\times 10^{-2}\,{\rm m\,s}^{-1}$&$-3/8$&$1/8$&$0$&$-D/2$\\
Diffusive collision rate&${\cal R}_{{\rm d}}$&eq.(\ref{eq: 6.3})&$1.8\times10^{-7}\,{\rm s}^{-1}$&$-9/4$&$3/4$&$-2$&$(4-3D)/2$\\
Advective collision rate&${\cal R}_{{\rm a}}$&eq.(\ref{eq: 6.8})&$1.4\times 10^{-15}\,{\rm s}^{-1}$&$-15/4$&$1$&$-2$&$2-D$\\
 Relative velocity&$\sqrt{\langle \Delta v^2\rangle}$&eq.(\ref{eq: 3.5}), $K\!=\!1$&$0.98\,{\rm ms}^{-1}$&$0$&$-1/4$&$2$&$(D-2)/2$\\
Gas-kinetic collision rate&${\cal R}_{{\rm g}}$&eq.(\ref{eq: 6.4})&$3.6\times 10^{-6}\,{\rm s}^{-1}$ &$-27/16$&$3/16$&$1$&$(2-D)/2$\\
Stokes number&${\rm St}$&eq.(\ref{eq: 6.5})&$0.17$&$3/8$&$-3/8$&$2$&$D-2$\\
\enddata
\end{deluxetable}

\begin{deluxetable}{lcccccc}
 \footnotesize
 \rotate
 \tablewidth{18cm}
 \tablecaption{\label{tab:3} Estimates for the compact particles
 ($D=3$) with ${\rm St}=1$. The results are of the form (\protect\ref{eq: 5.11}).}
 \tablehead{Variable&Symbol&Equation&$X(R_0)$&$\delta_R$&$\delta_\Lambda$&$\delta_\ell$}
 \startdata
 Stokes-trapped particle size&$a^\ast$&eq.(\ref{eq: 6.7})&$5.7\times 10^{-7}\,{\rm m}$&$-3/8$&$3/8$&$-2$\\
 Diffusive collision rate in Stokes-trap&${\cal R}_{{\rm d}}^\ast$ &eq.(\ref{eq: 6.3})&$2.3\times10^{-9}\,{\rm s}^{-1}$ &$-21/16$&$3/16$&$3$\\
   Advective collision rate in Stokes trap&${\cal R}_{{\rm a}}^\ast$&eq.(\ref{eq: 6.8}) &$2.5\times 10^{-13}\,{\rm s}^{-1}$&$-39/8$&$17/8$&$-8$\\
    Gas-kinetic collision rate in Stokes trap&${\cal R}_{{\rm g}}^\ast$&eq.(\ref{eq: 6.4}) & $3.6\times 10^{-6}\,{\rm s}^{-1}$& $-27/16$&$3/16$&$1$\\
   Turbulent relative velocity at ${\rm St}=1$&$\langle \Delta v\rangle^\ast$ &$({\cal E}\tau)^{1/2}$&$2.3\,{\rm m\,s}^{-1}$&$-3/16$&$-1/16$&$1$\\
  \enddata
\end{deluxetable}

The growth of the aggregates due to collisions could be be limited
by two factors. Firstly, the rate of collisions usually decreases
as the size of the aggregates increases and the collision rate
could become negligible when the particles grow to a sufficiently
large size. Secondly, large aggregates could be more vulnerable to
being fragmented upon collision.

\begin{figure}[t]
{\includegraphics[width=5cm]{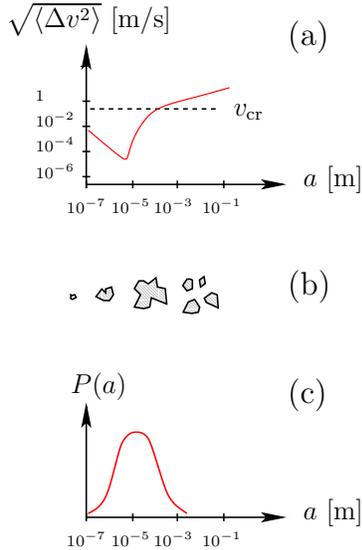}} \caption{\label{fig: 3}
Schematic illustration of the Stokes trap. (a): Relative collision
velocity of equal-sized aggregates versus their size $a$. (b):
Small aggregates grow by collisions, but larger aggregates tend to
fragment upon collision. (c): Corresponding stable particle-size
distribution.}
\end{figure}

To illuminate a discussion of whether these limitations occur in
practice, we now consider the properties of particles with size
$a$ chosen so that ${\rm St}=1$. For compact particles ($D=3$) we
find that their size is
\begin{equation}
\label{eq: 6.7} a^\ast=a_0\left(\frac{1}{{\rm
St}_0}\right)^{1\over{D-2}}\sim
a^\ast(R_0)\Lambda^{3/8}\,\ell^{-2}\left(\frac{R}{R_0}\right)^{-3/8}
\end{equation}
with $a^\ast(R_0)\approx 6\times 10^{-7}{\rm m}$: for $\ell=0.1$
this leads to particles of size $a^\ast\approx 25\,\mu{\rm m}$ at
$R=10\,{\rm AU}$. It is also of interest to evaluate the collision
rate of these particles. Below we describe a mechanism which will
lead to the aggregates becoming compact, so that we concentrate on
the predictions of this calculation when $D=3$. The approximate
values are listed in table \ref{tab:3}. Note that the diffusive
collision rate is very small, but it increases abruptly (by
approximately five orders of magnitude, for the case where
$\ell=0.1$) when the size of the particles increases such that the
Stokes number exceeds unity (see figure~\ref{fig: 3}(a)). Also,
note that the relative velocity for collisions of particles with
Stokes number unity is comparable to the critical velocity for
microscopic dust aggregates to be fragmented.

For compact aggregates ($D\approx 3$), the collision rate for the
gas-kinetic mechanism is the much larger when ${\rm St}\gg 1$, and
it also decreases less rapidly as the size of the particles
increases. If the lifetime of the accretion system is $t_{\rm
acc}\approx 10^6{\rm yr}$, the growth process may be assumed to be
ended when ${\cal R}_{\rm g}t_{\rm acc}=1$. The estimate for
${\cal R}_{\rm g}$ above indicates that the gas-kinetic collision
mechanism is sufficiently rapid that it does not limit the growth
of aggregates, at least until they are so large that the
approximations in equation (\ref{eq: 6.5}), such as assuming that
the drag is determined by the Epstein formula, break down. We
conclude that kinetic factors do not limit the growth of
aggregates, once particles become sufficiently large that ${\rm
St}>1$. Note, however that for compact particles at the size where
${\rm St}=1$, the diffusive collision rate is extremely slow (of
the order of one collision per $10^5$ years when $\ell=0.1$). The
slowing down of the collision rate as the particle size increases
is therefore quite close to becoming an insurmountable bottleneck.

As the size $a$ of the aggregates increases, so does their
relative velocity $\langle \Delta v\rangle$, for example compact
aggregates of size $a=10\,{\rm cm}$ collide with a relative
velocity of approximately $10\,{\rm ms}^{-1}$ if we take
$\ell=0.1$. Such a collision would certainly cause the aggregates
to fragment upon impact.

We have arrived at the following picture of the initial stages of
the growth of dust particle aggregates. Initially, dust particles
will aggregate by Brownian diffusion. This process continues until
the Stokes number of the larger particles is of order unity. Then
the velocity of these particles starts to separate from that of
the gas due to the formation of fold caustics. This greatly
increases the relative velocity between these particles and
smaller ones in their vicinity (figure~\ref{fig: 3}(a)): the
larger particles will rapidly \lq sweep up' the remaining small
particles in their vicinity. However, when these larger particles
collide with each other they have sufficient kinetic energy that
they cause each other to fragment, so that particles with large
Stokes number fragment to replenish the supply of particles with
small Stokes number (figure~\ref{fig: 3}(b)). This results in a
steady-state distribution of particle sizes, corresponding to
Stokes numbers of order unity (illustrated in figure~\ref{fig:
3}(c)): we call this the \lq Stokes trap'.

\section{Mechanism for avoiding the Stokes trap}
\label{sec: 7}

In the following we propose four mechanisms by which the Stokes
trap might be avoided.

\subsection{Escaping the Stokes trap by clustering}
\label{sec: 7b}

It is known that small particles suspended in a turbulent flow can
cluster, even if the fluid flow is incompressible, due to effects
of the inertia of the particles. This effect was proposed by
\citet{Max87}, who suggested that heavy particles would be \lq
centrifuged' out of vortices and would tend to cluster in regions
of low vorticity: this effect is often referred to as \lq
preferential concentration' because it is thought that the
particles cluster in regions of low vorticity.

The argument by Maxey does not give a quantitative description of
the nature of the clustering effect at long times.  \citet{Som93}
suggested that particles in random fluid flows cluster onto a
fractal set of dimension $D_{\rm cl}$ and proposed using the
Lyapunov dimension (first defined by \citet{Kap79}) to
characterise this set. Recent numerical experiments \citep{Bec06}
have studied how the (Lyapunov) fractal dimension $D_{\rm L}$ of
particles in a three-dimensional turbulent flow varies as a
function of Stokes number: the minimum value is $D_{\rm L}\approx
2.6$ for ${\rm St}=0.55$, $D_{\rm L}=3$ for ${\rm St}>{\rm
St}_{\rm c}\approx 1.7$ and $D_{\rm L}\to 3$ as ${\rm St}\to 0$.
Theoretical work \citep{Dun05} has shown that the fractal
clustering may be explained without the centrifuge (or \lq
preferential concentration') effect: a model for which it is
absent gives a minimum dimension $D_{\rm L}\approx 2.65$, in good
agreement with simulations.

It has previously been suggested that this clustering effect may
play a role in planet formation \citep{Cuz01}, but there is a
difficulty. The clustering effect only occurs when ${\rm
St}=O(1)$, but most aggregation processes produce a broad (and
time-dependent) distribution of sizes. The clustering mechanism
would then only act on a small proportion of the particles. Our
Stokes-trapping mechanism avoids this difficulty.

The Stokes-trapped dust aggregates will be clustered, but
if the clustering is confined to length scales below
the Kolmogorov length $\eta$,
the effect is not sufficiently strong to initiate gravitational
collapse:  the mass of solid matter within a Kolmogorov length is
\begin{equation}
\label{eq: 7b.1} M\sim \kappa \rho_{\rm g}\eta^3\approx 1.0\times
10^{-3}\,{\rm
kg}\,\Lambda^{-11/16}\,\ell\left(\frac{R}{R_0}\right)^{51/16}\ .
\end{equation}
The size of the clusters created by the \lq preferential
concentration' effect cannot significantly exceed this value, and
this mass is much too small for gravitational effects to become
significant.

We thus conclude that although the clustering effect is promoted by the
Stokes trap, clustering of dust aggregates onto a fractal set
cannot provide inhomogeneities which are sufficient to trigger
gravitational instability in protoplanetary systems. However, it is
possible that this clustering effect could influence the
scattering of electromagnetic radiation by the dust grains.

Some other mechanisms for clustering have been proposed, discussed in
\citet{Bar95}, \citet{Joh07}. We argue that these mechanisms can be discounted
because they assume the existence of much larger
aggregates than the Stokes-trapped size.

\subsection{Escaping the Stokes trap by chemical processes}
\label{sec: 7a} We propose that the mid-plane of the accretion disk can be at a
much higher temperature than its surface. The dust aggregates
circulate between regions with different temperatures. In the
lower temperature regions, water and organic compounds can condense onto
the aggregates. When the particles circulate to higher temperature regions
chemical reactions may occur. This is expected to both compactify
the aggregate and to make it much more resistant to being
fragmented in collisions. It is also possible that the temperature
in regions close to the star is sufficiently high that at least
some of their component materials are melted. This will also
produce much more robust and compact aggregates, able to withstand
collisions at much higher relative velocities.

We have assume that the temperature of the accretion disk is
determined by dissipative heating, and we have estimated the
surface temperatures, which are quite low. However, if the disk is
optically thick in the spectral region corresponding to the Wien
wavelength, the mid-plane temperature may be considerably higher
than the surface temperature.

The optical thickness $W$ is the ratio of the height of the disk
to the photon mean free path. The optical properties are dominated
by the dust particles if $W\gg 1$, and when the particles are
larger than the wavelength we may assume that their optical
cross-section $S_{\rm opt}$ is approximately the square of their
linear dimension: $S_{\rm opt}\approx a^2$. If the density of
particles is $n$, the optical mean-free path is $\lambda_{\rm
opt}\sim 1/(nS_{\rm opt})$. Thus for compact particles ($D=3$),
the optical thickness is
\begin{equation}
\label{eq: 7a.1} W\sim Hna^2\ .
\end{equation}
The ratio of the interior temperature $T_{\rm int}$ of the
accretion disk to its surface temperature $T$ is
\begin{equation}
\label{eq: 7a.2} \frac{T_{\rm int}}{T}\sim W^{1/4}\ .
\end{equation}
Very small particles, with dimension small compared to the Wien
wavelength corresponding to temperature $T$, do not absorb or
scatter radiation effectively. If the aggregates are very large,
their number is correspondingly reduced, so that they make a
smaller contribution to the optical thickness. From the data
above, we see that the Stokes-trapped particles have sizes which
are comparable to, but somewhat larger than, the Wien wavelength
corresponding to the surface temperature. The Stokes-trapped
particles are therefore of approximately optimal size to increase
the optical thickness. The optical thickness corresponding to
Stokes-trapped particles is therefore (for compact particles,
$D=3$):
\begin{equation}
\label{eq: 7a.3} W^\ast\sim \frac{H n_0 a_0^3}{a^\ast}\sim 410\,
\Lambda^{3/8}\left(\frac{R}{R_0}\right)^{-3/8}\ .
\end{equation}
The temperature at the centre of the disk is therefore expected to exceed the
surface temperature by a factor of approximately four in the range
$R=1-10{\rm AU}$.

The chondrules, grains of typically submillimetre size which are
found in many meteorites, show evidence of having been heated to
very high temperatures, causing melting. The temperatures required
are rather higher than those which we predict above (except very
close to the star). Also, the exposure of chondrules to high
temperatures may have been brief compared to the circulation time
$t_L$, otherwise the chondrules may have evaporated.

\subsection{Escaping the Stokes trap by forming tenuous aggregates}
\label{sec: 7c}

We have already remarked that if the fractal
dimension $D$ of the dust aggregates is $D\le 2$, then the Stokes
number does not increase as the size of the aggregate increases.
Such aggregates would avoid the \lq Stokes trap', but there are
three reasons why these tenuous aggregates might not provide a
satisfactory solution to the problem of planet formation.

First, it might be thought that as such aggregates grow their
collision rate would become so small that growth would, for all
practical purposes, cease. In section \ref{sec: 4} it was
determined how the collision rate for fractal aggregates with
dimension $D$ scales as a function of their characteristic size,
$a$: we found ${\cal R}_{\rm d} \propto  a^{(4-3D)/{2}}$.
The collision rate therefore increases with the aggregate size
when $D<4/3$, because such very tenuous aggregates are so extended
that they entangle each other. Experiments on growing aggregates
in negligible gravity have produced objects with very low fractal
dimensions of roughly $D=1.4$ \citep{Blu00,Kra04}, rather smaller
than the values predicted for kinetic aggregation processes. The
dust aggregates observed in low-gravity experiments are only
observed for a short period compared to the time between
collisions between dust aggregates, so that the structure of the
aggregates in protostellar accretion systems is uncertain, but
there is probably no kinetic restriction on the growth of very
tenuous clusters.

A second possibility is that the tenuous aggregates are torn apart
by the effect of the shearing motion of the fluid. We estimate
this effect as follows. Consider the force required to split an
aggregate of fractal dimension $D$, size $a$, composed of
particles of size $a_0$. The number of bonds at an equatorial
plane of the aggregate is approximately
\begin{equation}
\label{eq: 7c.2} N_{\rm b}\sim \left(\frac{a}{a_0}\right)^{D-1}\ .
\end{equation}
Each bond has a breaking strain $F_{\rm b}$. Grains in contact
have an energy per unit area $\delta E_1/l^2$ per unit area, acting
over a range $l$. The binding energy is $E_{\rm b}=\pi\delta
E_1a_0/l$ (see section \ref{sec: 4.3}). Writing $E_{\rm b}=F_{\rm b}l$, we estimate
\begin{equation}
\label{eq: 7c.3} F_{\rm b}\sim \frac{\delta E_1}{l^2}a_0\ .
\end{equation}
The drag force on a single grain due to fluid motion with a speed
$u$ relative to the fluid is $F=m_0\gamma_0 u$. When $D\le 2$, the
drag force pulling the aggregate apart is comparable to the sum of
the forces which are predicted by applying this formula to each
grain. The relative velocity of the gas at opposite sides of the
aggregate is $a/\tau$, so that the force acting to break up the
aggregate is
\begin{equation}
\label{eq: 7c.4} F_{\rm br}\sim \frac{N\gamma_0m_0a}{\tau}\sim
\frac{m_0\gamma_0 a_0}{\tau}\left(\frac{a}{a_0}\right)^{D+1}\ .
\end{equation}
The ratio of the applied force to the critical force is $F_{\rm
br}/N_{\rm b}F_{\rm b}$; note that this ratio is independent of
$D$ (provided $D\le 2$). Setting this ratio equal to unity we
estimate the size $a_{\rm max}$ of aggregates which will be torn
apart by shearing forces:
\begin{equation}
\label{eq: 7c.5} a_{\rm max}\approx 78\,{\rm
m}\,\Lambda^{-9/16}\,\ell\left(\frac{R}{R_0}\right)^{33/16}\ .
\end{equation}
The estimates leading to this result are valid provided the
aggregate is small compared to the Kolmogorov length: $a\ll \eta$:
this condition is only marginally satisfied. It is not clear how
the shear forces increase with aggregate size once the aggregate
is large compared to the Kolmogorov correlation scale, but it is
hard to escape the conclusion that tenuous aggregates are torn
apart by shear forces when they become sufficiently large. The
mass contained in these tenuous aggregates is clearly comparable
to that estimated in equation (\ref{eq: 7b.1}) and is too small to
initiate gravitational collapse directly.

A third problem is that aggregates with Stokes number less than
unity are always advected with the gas. Even if they could become
very massive, they would never collapse to the mid-plane of the
disk, as long as the motion of the gas remains turbulent.

\subsection{Escaping the Stokes trap by reducing the accretion rate}
\label{sec: 7d}

If the accretion rate $\dot M$ is reduced (by choosing a small
value for the parameter $\Lambda$ in equation (\ref{eq: 5.0})),
there is a corresponding reduction in the turbulence intensity
${\cal E}$. We might expect that there would be also be a
reduction in the relative velocities of suspended particles. This
latter expectation is false, because a reduction in the rate of
accretion results in a reduced density of the gas, and the
suspended particles are more lightly damped by the gas (note that
if $\gamma$ decreases the relative velocity increases: see
equation (\ref{eq: 3.5})).

This principle is illustrated by results in table \ref{tab:3},
which show that the Stokes-trapped particle size decreases as the
accretion rate parameter $\Lambda$ decreases (because
$\delta_\Lambda>0$ for $a^\ast$). A further demonstration comes
from considering the size of particles which are not fragmented on
collision. If the relative velocity is less than the critical
velocity $v_{\rm cr}$ for disintegration of dust aggregates, then
they continue to grow on collision. The size of particles for
which the turbulent relative velocity exceeds $v_{\rm cr}$ is
\begin{equation}
\label{eq: 7d.1} a_{\rm cr}=1.0\times 10^{-7}\ {\rm s}^2{\rm
m}^{-1}v_{\rm cr}^2\,\ell^{-4}\,\Lambda^{1/2}\ .
\end{equation}
We see that, contrary to intuition, the attainable particle size
decreases as the rate of accretion decreases.

\section{Settling and gravitational collapse}
\label{sec: 8}

Even if the Stokes trap is circumvented, the particles must grow
to a substantial size in order for them to settle to the mid-plane
of the accretion disk. In the following we illustrate this fact by
estimating the size at which particles start to settle to the
mid-plane, according to our model. We find that settling only
starts at large particle sizes (approximately $10\,{\rm cm}$,
where the use of the Epstein formula for the damping rate becomes
questionable). We also estimate the critical height of the dust
layer for the onset of gravitational collapse, and find that this
is a small fraction of the height of the gas layer.

\subsection{Critical size for settling}
\label{sec: 8.1}

The integral scale fluctuations with velocity $u_L$ will advect
aggregates away from the mid plane. This effect is opposed by
gravitational attraction to the mid plane. The gravitational
acceleration at a distance $z$ from the mid plane is
\begin{equation}
\label{eq: 8.1} g(R,z)={GM\over R^3}z=\Omega^2z\ .
\end{equation}
If the integral timescale of the turbulence, $t_L$, is large
compared the the relaxation time $\gamma^{-1}$, then the
aggregates only settle to the mid-plane if their terminal speed
$u_{\rm t}=g(R,z)/\gamma$ exceeds $u_L$. There is thus no settling
at all unless $\gamma<\gamma_{\rm s}$, where
\begin{equation}
\label{eq: 8.2} \gamma_{\rm s}={\Omega^2H\over{u_L}}=1.8\times
10^{-7}\,{\rm s}^{-1}\ell^{-1}\left(\frac{R}{R_0}\right)^{-3/2}\ .
\end{equation}
In the case of compact aggregates ($D=3$), we conclude that
particles start to settle to the mid-plane when their size exceeds
\begin{equation}
\label{eq: 8.3} a_{\rm s}\approx
a_0\left(\frac{\gamma_0}{\gamma_{\rm s}}\right)\approx 6.0\times
10^{-2}{\rm
m}\,\Lambda^{3/4}\,\ell^{-1}\left(\frac{R}{R_0}\right)^{-3/4}\ .
\end{equation}
When considering very large aggregates, the damping rate $\gamma $ may be
sufficiently small that $\gamma t_L\ll 1$. In this case, the effect of the
turbulence must be modelled as a sequence of random impulses on the aggregate
\citep{Cuz93,You07}. We argue that, because of the
Stokes trapping effect, the aggregates never grow to a sufficient size for this
alternative approach to become relevant.

\subsection{Critical height for gravitational collapse}
\label{sec 8.2}

A stationary gas of solid particles with mass density $\rho_{\rm
sol}$ is unstable against gravitational collapse on a timescale
$t_{\rm coll}\sim (G\rho_{\rm sol})^{-1/2}$. If the gas consists
of particles with typical relative speed $\Delta v$, then
gravitational collapse (the Jeans instability) occurs on length
scales greater than $\Delta v t_{\rm coll}$. In the case of an
accretion disk, the question of gravitational instability can be
complicated by the rotational motion and the finite thickness of
the disk.

The gravitational stability of a uniform thin disk of area density
$\Sigma$ may be described by giving a dispersion relation for the
frequency $\omega$ of a density perturbation with wavenumber $k$
\begin{equation}
\label{eq: 8.4} \omega^2=\Omega^2-2\pi G\Sigma \vert k\vert+v_{\rm
s}k^2
\end{equation}
where $v_{\rm s}$ is a two-dimensional sound velocity \citep{Bin88}. The
system becomes gravitationally unstable when $\omega^2<0$ for any
value of $k$. For a collisionless system, we have $v_{\rm s}=0$,
and the instability arises when the term containing the
gravitational constant is larger in magnitude that the centrifugal
term. The instability is favoured by choosing a large value for
$k$, but if we consider very large values of $k$, the
approximation of treating the mass distribution as two-dimensional
fails. We therefore assume that the largest possible value of $k$
is $2\pi/H$, where $H$ is the height of the disk.

The solid material in the disk (with area density $\kappa \Sigma$)
therefore becomes gravitationally unstable when its scale height,
$H_{\rm sol}$ is less than a critical value which is given
(approximately) by
\begin{equation}
\label{eq: 8.5} H_{\rm cr}=\frac{4\pi^2 \kappa
G\Sigma}{\Omega^2}\approx 3.3\times 10^5{\rm
m}\,\Lambda^{3/4}\,\ell^{-2}\left(\frac{R}{R_0}\right)^{9/4}\ .
\end{equation}
Note that $H_{\rm cr}/H\ll 1$, implying that to trigger a
gravitational instability, particles must be very much heavier than
particles of the size $a_{\rm cr}$ (given by equation (\ref{eq:
8.3})).

\section{Concluding remarks}
\label{sec: 9}

In this paper we have analysed the consequences of turbulence for
a standard model of the formation of planets by aggregation of
sub-micron sized dust. This led us to introduce the concept of the
Stokes trap, a mechanism which limits the growth of dust
aggregates in turbulent protostellar accretion disks and thus
constitutes a barrier to planet formation. The particle sizes
predicted by the Stokes trap mechanism are so small that turbulent
fluctuations would never allow the particles to settle to the
mid-plane of the disk and achieve sufficient density to trigger
gravitational instability. The Stokes trap must therefore be
avoided if the planets are to form.

We have proposed four possible ways of avoiding the Stokes trap
and assessed these mechanisms in the light of Kolomogorov's
scaling principle for turbulence in combination with a standard
model for the accretion disk. Particles clustering due to \lq
preferential concentration' cannot produce sufficiently large
clusters. Chemical processes which consolidate aggregates will
strengthen them, but not to the extent that large aggregates would
be invulnerable to fragmentation by collision. If the aggregates
are very tenuous fractals, with dimension $D\le 2$, their Stokes
number remains very small, but the aggregates may be torn apart by
shearing forces when they become sufficiently large, and they will
not settle to the mid-plane of the disk. Finally, we observed that
as the disk becomes less active (that is, as $\dot M$ decreases)
the relative velocity of the suspended aggregates increases, so
that the difficulty is not removed as the disk becomes quiescent.

Even if the aggregates escape from the Stokes trap and continue
growing, our estimates indicate that they must reach very large
sizes before they can overcome the effects of turbulence and start
to settle to the mid-plane. Our estimates indicate that the
turbulence intensity would have to be reduced by many orders of magnitude
before this could happen.

We conclude that a model for planet growing by aggregation of
small dust grains is highly problematic. It is therefore necessary
to examine other possibilities. The difficulties with the theory
are avoided if the protoplanetary accretion disk already contains
large objects with a gravitational escape velocity which is large
enough to prevent them from being disrupted by collisions with
dust grains and aggregates. A star forms when a cloud of
interstellar gas satisfies conditions which allow it to undergo
gravitational collapse. It is likely that at the same time as the
star forms, other regions of the cloud in the vicinity of the
nascent star will condense by gravitational attraction. Condensed
objects much smaller than the star itself may be drawn into the
accretion disk of the star, and eventually become planets lying on
roughly circular orbits in the equatorial plane of the star. Thus
we propose an alternative picture for the formation of planets
which we term \lq concurrent collapse', in which the planets are
formed as gravitationally bound systems at the same time as the
star. During the life of the accretion disk they become coupled by
friction and mass transfer to the circular motion of the accretion
disk. While in the accretion disk their structure may be radically
transformed by processes such as further accretion of material.
Their mass might also be reduced by ablation or evaporation of
lighter elements.

One possible criticism of this scenario is that the density $\rho$
of the gas cloud at temperature $T$ which collapses to form the
star will be associated with a mass scale, the Jeans mass,
\begin{equation}
\label{eq: 9.1} M_{\rm J}\sim
\left[\frac{1}{\rho}\left(\frac{k_{\rm B}T}{G m_{\rm
H}}\right)^3\right]^{1/2}\ .
\end{equation}
It may be argued that the Jeans mass is a lower limit to the mass of objects that could form by
gravitational collapse. However, as gravitational collapse proceeds, the density of some regions
will increase, thus lowering the Jeans mass estimate which applies for further collapse. Images of
gas clouds suggest that they are typically non-uniform in their
density, and that they have density fluctuations spanning a wide range of spatial scales. The distribution
of sizes of dense objects produced by gravitational collapse from such a non-uniform initial condition
might be expected to be very broad.

For the time being our \lq concurrent collapse' hypothesis must be supported by the
implausibility of the standard dust aggregation model. Because the additional hypothesis only concerns
conditions at early stages of the life of the stellar accretion system, its implications for
the eventual structure of a solar system are unclear.

\acknowledgements
Support from Vetenskapsr\aa{}det and
the platform \lq Nanoparticles in an interactive environment'
at G\"oteborg university are gratefully acknowledged.


\begin{thebibliography}{}

\bibitem[Abrahamson(1975)]{Abr75} J. Abrahamson,
Collision rates of small particles in a vigorously turbulent fluid,
{\it Chem. Eng. Sci.}, {\bf 30}, 1371-9, (1975)

\bibitem[Apai {\em et al.}(2005)]{Apa05} D. Apai {\em et al.},
The onset of planet formation in brown dwarf disks, {\it  Science}, {\bf 310}, 834, (2005)

\bibitem[Armitage(2007)]{Arm07} P. J. Armitage,
Lecture notes on the formation and early evolution of planetary
systems, arXiv:astro-ph/0701485

\bibitem[Barge \& Sommeria(1995)]{Bar95}
P. Barge and J. Sommeria, Did planet formation begin inside persistent gaseous vortices?,
{\it Astron. Astrophys.}, {\bf 295}, L1-4, (1995).

\bibitem[Bauer {\em et al.}(2007)]{Bau07}
F. Bauer, C. P. Dullemond, A. Johansen, Th. Henning, H. Klahr and
A. Natta, Survival of the mm-cm size grain population observed in
protoplanetary disks, arXiv:astro-ph/0704.2332

\bibitem[Bec {\em et al.}(2006)]{Bec06}
J. Bec, L. Biferale, G. Boffetta, M. Cencini, S. Musachchio and F.
Toschi, Lyapunov exponents of heavy particles in turbulence, {\it
Phys. Fluids}, {\bf 18}, 091702, (2006)

\bibitem[Binney \& Tremaine(1988)]{Bin88} J. Binney and S. Tremaine, {\it Galactic Dynamics},
Princeton University Press, (1988)

\bibitem[Blum {\em et al.}(2000)]{Blu00} J. Blum {\it et al},
Growth and form of planteary seedlings: results from a
microgravity aggregation experiment, {\it Phys. Rev. Lett.}, {\bf
85}, 2426-9, (2000)

\bibitem[Carciofi {\em et al.}(2004)]{Car04} A. C. Carciofi,
J. E. Bjorkman and A. M. Magalhaes, Effects of grain size of the
spectral energy distribution of dusty circumstellar envelopes,
{\it Astrophysical J.}, {\bf 604}, 238-51, (2004)

\bibitem[Chokshi {\em et al.}(1993)]{Cho93}
A. Chokshi, A. G. G. M. Tielens and D. Hollenbach, Dust coagulation,
{\it Astrophys. J.}, {\bf 407}, 806-819, (1993)

\bibitem[Cuzzi {\em et al}(1993)]{Cuz93}
J. N. Cuzzi, A. R. Dobrovolskis and J. M. Champney, Particle-gas
dynamics in the mid-plane of a protoplanetary nebula, {\it
Icarus}, {\bf 106}, 102-34, (1993)

\bibitem[Cuzzi {\em et al.}(2001)]{Cuz01}
J. N. Cuzzi, R. C. Hogan, J. M. Paque, A. R. Dobrovolskis,
Size-selective concentration of chondrules and other small particles in protoplanetary nebula turbulence,
{\it Astrohys. J.}, {\bf 546}, 469-508, (2001)

\bibitem[Dominik {\em et al.}(2007)]{Dom07} C. Dominik, J. Blum, J. N. Cuzzi and G. Wurm,
Growth of dust as the initial step toward planet formation, {\it
Protostars and protoplanets V}, B. Reipurth, D. Jewitt and K. Keil
(eds.), 783-800, (2007)

\bibitem[Dominik \& Tielens(1997)]{Dom97}
C. Dominik and A. G. G. M. Tielens, The physics of dust
coagulation and the structure of dust aggregates in space, {\it
Astrophys. J.}, {\bf 480}, 647-73, (1997)

\bibitem[Duncan {\em et al.}(2005)]{Dun05}
K. P. Duncan, B. Mehlig, S. \" Ostlund and M. Wilkinson,
Clustering in mixing flows, {\it Phys. Rev. Lett.}, {\bf 95},
240602, (2005)

\bibitem[Epstein (1924)]{Eps24} P. S. Epstein,
On the resistance experienced by spheres moving through gases,
{\it Phys. Rev.}, {\bf 23}, 710, (1924)

\bibitem[Falkovich {\em et al.}(2002)]{Fal02} G. Falkovich, A. Fouxon and G. Stepanov,
Acceleration of rain initiation by cloud turbulence, {\it Nature}, {\bf 419}, 151-154, (2002)

\bibitem[Falkovich \& Pumir(2006)]{Pum06} G. Falkovich and A. Pumir,
Sling effect in collisions of water droplets in clouds,
arXiv:nlin/0605040, (2006)

\bibitem[Frisch(1997)]{Fri97} U. Frisch, {\it Turbulence}, Cambridge University Press, (1997)

\bibitem[Hartmann {\em et al.}(1998)]{Har98} L. Hartmann, N. Calvet, E. Gullbring and P D'Aessio,
Accretion and evolution of T-Tauri disks, {\it Astrophysical J.},
{\bf 495}, 385, (1998)

\bibitem[Hayashi(1981)]{Hay81} C. Hayashi,
Structure of the solar nebula, growth and decay of magnetic fields
and effects of magnetic and turbulent viscosities of the nebula,
{\it Suppl. Prog. Theor. Phys.}, {\bf 70}, 35-53, (1981)

\bibitem[Henning {\em et al.}(2006)]{Hen06} T. Henning, C. P. Dullemond, S. Wolf,
Dust coagulation in protoplanetary disks, in {\it Planet
Formation}, ed. H. Klar and W. Brandner, in press, (2006)

\bibitem[Johansen {\em et al}(2007)]{Joh07}
A. Johansen, J. S. Oishi, M-M. M. Low, H. Klahr, T. Henning and A.
Youdin, Rapid Planetesimal Formation in Turbulent Circumstellar
Discs, {\it Nature}, {\bf 448}, 1022-5, (2007)

\bibitem[Kaplan \& Yorke(1979)]{Kap79}
J. L. Kaplan and J. A. Yorke, in: {\it Functional differential
equations and approximations of fixed points}, eds.: H.-O. Peitgen
and H.-O. Walter, Lecture notes in mathematics, {\bf 730},
Springer, Berlin, (1979), p. 204

\bibitem[Kessler-Silacci {\em et al.}(2007)]{Kes07} J. E. Kessler-Silacci, C. P. Dullemond and J. C. Augereau,
Probing protoplanetary disks with silicate emission: where is the
silicate emission zone?, {\it Astrophysical J.}, {\bf 659}, 680-4,
(2007)

\bibitem[Krause \& Blum(2004)]{Kra04} M. Krause and J. Blum,
Growth and form of planteary seedlings: results from a sounding
rocket microgravity aggregation experiment, {\it Phys. Rev.
Lett.}, {\bf 93}, 021103, (2004)

\bibitem[Markiewicz {\em et al}(1991)]{Mar91}
W. J. Markiewicz, H. Mizuno and H. J. V\"olk, Turbulence induced
relative velocity between two grains, {\it Astron. Astropkys.},
{\bf 242}, 286, (1991)

\bibitem[Maxey(1987)]{Max87} M. R. Maxey, The gravitational settling
of aerosol paerticles in homogeneous turbulence and random flow
field, {\it J. Fluid Mech.}, {\bf 174}, 441, (1987)

\bibitem[Mehlig {\em et al.}(2007)]{Meh07} B. Mehlig, V. Uski and M. Wilkinson,
Colliding particles in highly turbulent flows, {\it Phys. Fluids},
{\bf 19}, 098107, (2007)

\bibitem[Mizuno {\em et al.}(1988)]{Miz88} H. Mizuno, W. J. Markiewicz and H. J. V\"olk,
Grain growth in turbulent protoplanetary accretion disks,
{\it Astron. Astrophys.}, {\bf 195}, 183-92, (1988)

\bibitem[Nakamura {\em et al.}(1994)]{Nak94} R. Nakamura, Y. Kitada and T. Mukai,
Gas drag forces on fractal aggregates, {\it Planet. Space Sci.}, {\bf 42}, 771-6, (1994)

\bibitem[Poppe \& Blum(1997)]{Pop97}
T. Poppe and J. Blum, Experiments on pre-planetary grain growth, {\it Adv. Space Res.}, {\bf 20},
1595-1604, (1997)

\bibitem[Saffman \& Turner(1956)]{Saf56} P. G. Saffman and J. S. Turner,
On the collision of drops in turbulent clouds, {\it J. Fluid Mech.}, {\bf 1}, 16-30, (1956)

\bibitem[Safranov(1969)]{Saf69}
V. S. Safranov, Evoliutsiia Doplanetnogo Oblaka, (1969) (English transl.: Evolution
of the protoplanetary cloud and formation of the Earth and planets, NASA Tech. Transl. F-677,
Jerusalem: Israel Sci. Transl., 1972)

\bibitem[Shakura \& Sunyaev(1973)]{Sha73} N. I. Shakura and R. A. Sunyaev,
Black holes in binary systems. Observational appearance, {\it
Astronomy and Astrophysics}, {\bf 24}, 337, (1973)

\bibitem[Sommerer \& Ott(1993)]{Som93} J. Sommerer and E. Ott,
Particles floating on a fluid - a dynamically comprehensible physical fractal,
{\it Science}, {\bf 259}, 335, (1993)

\bibitem[Sundaram \& Collins(1997)]{Sun97} S. Sundaram and L. R. Collins,
Collision statistics in a isotropic particle-laden turbulent suspension.  Part 1. Direct numerical simulations,
{\it J. Fluid Mech.}, {\bf 335}, 75-109, (1997)

\bibitem[van Boekel {\em et al.}(2003)]{vBo03} R. van Boekel {\em et al.},
Grain growth in the inner regions of Herbig Ae/Be star disks, {\it
Astron. \& Astrophys.}, {\bf 400}, L21, (2003)

\bibitem[V\"olk {\em et al.}(1980)]{Vol80} H. J. V\"olk, F. C. Jones, G. E. Morfill and S.  R\"oser,
Collisions between grains in a turbulent gas, {\it Astron. \&
Astrophys.}, {\bf 85}, 316, (1980)

\bibitem[Wilkinson \& Mehlig(2005)]{Wil05} M. Wilkinson and B. Mehlig,
Caustics in turbulent aerosols, {\it Europhysics Lett.}, {\bf 71}, 186-92, (2005)

\bibitem[Wilkinson {\em et al.}(2006)]{Wil06} M. Wilkinson, B. Mehlig and V. Bezuglyy,
Caustic activation of rain showers, {\it Phys. Rev. Lett.}, {\bf
97}, 048501, (2006)

\bibitem[Wood {\em et al.}(2002)]{Woo02} K. Wood, M. J. Wolff, J. E. Bjorkman and B. Whitney,
The spectral energy distribution of HH 30 IRS: constraining the
circumstellar dust size distribution, {\it Astrophysical J.}, {\bf
564}, 887, (2002)

\bibitem[Wurm \& Blum(1998)]{Wur98}
G Wurm and J. Blum, Experiments on preplanetary dust aggregation,
{\it Icarus}, {\bf 132}, 125-36, (1998)

\bibitem[Youdin(2003)]{You03} A. N. Youdin,
Obstacles to the collisional growth of planetesimals,
arXiv:astro-ph/0311191

\bibitem[Youdin(2007)]{You07}
A. N. Youdin and Y. Lithwick, Particle stirring in turbulent gas
disks: including orbital oscillations, arXiv:astro-ph/0707.2975

\end{thebibliography}
\end{document}